\documentclass[aps,superscriptaddress,preprint,showpacs]{revtex4}
\usepackage{amsmath}
\usepackage{graphicx}

\begin{document}

\title{On description of quantum plasma}
\author{S.~V.~Vladimirov}
\affiliation{School of Physics, The University of Sydney, NSW 2006, Australia}
\affiliation{Joint Institute of High Temperatures RAS, 125412 Moscow, Russian Federation}
\author{Yu.~O.~Tyshetskiy}
\affiliation{School of Physics, The University of Sydney, NSW 2006, Australia}
\affiliation{Department of Physics and Technology, Kharkiv National University, Ukraine}

\date{\today}
\received{}

\begin{abstract}
A plasma becomes quantum when the quantum nature of its particles significantly affects its macroscopic properties. To answer the question of when the collective quantum plasma effects are important, a proper description of such effects is necessary. We consider here the most common methods of description of quantum plasma, along with the related assumptions and applicability limits. In particular, we analyze in detail the hydrodynamic description of quantum plasma, as well as discuss some kinetic features of analytic properties of linear dielectric response function in quantum plasma. We point out the most important, in our view, fundamental problems occurring already in the linear approximation and requiring further investigation. (\textit{Submitted to Physics-Uspekhi})
\end{abstract}
\pacs{05.30.Fk,52.35.Mw,52.35.Ra,52.35Sb}

\maketitle

\tableofcontents

\newpage

\section{Introduction \label{sec:1}}

Plasma can be regarded as quantum when the quantum nature of its particles significantly affects its macroscopic properties. To determine when quantum effects in plasmas are important, their adequate description is needed. As plasma is an ensemble of many particles, the corresponding approach must be based on the appropriate description of a quantum particle.

The description of non-relativistic quantum plasma can be based on either Schr\"odinger's representation (in which the operators are time-independent, while the time dependence of physical quantities of the system is defined by the corresponding time dependence of the system's wave function or density matrix) or Heisenberg's representation (in which the time dependence is transferred from the wave functions to the operators). Most of quantum plasma models that are commonly used now~\cite{Manfredi_Howto} use the Schr\"odinger's representation; the quantum plasma state is described either by wave functions of separate particles (the so-called multistream model~\cite{Haasetal2000}), or by the density matrix, or by the Wigner function defined in terms of the density matrix in coordinate representation~\cite{Klim_Silin_52,Tatarskii_1983}, or -- and this approach has become popular recently -- by a set of the so-called quantum hydrodynamics equations. Naturally, simplifying assumptions are made in all these models (which thus lead to limitations of their applicability), which one should take into account when analyzing results obtained from them. However, concrete applicability limits of results obtained from a particular model are not always stated explicitly (this especially concerns the widely used model of quantum hydrodynamics~\cite{Shukla_Eliasson_UFN_2010,Eliassonetal2008}), which can lead to their incorrect interpretation. This has recently been pointed out, for example, by Melrose and Mushtaq~\cite{Melrose_Mushtaq_2009} as well as by Kuzelev and Rukhadze~\cite{Kuzelev_Rukhadze_UFN_2010}.

With the recent rapid increase in the number of publications on quantum plasmas, the lack of detailed analysis of the made assumptions and the associated limitations for the most common quantum plasma models becomes increasingly obvious, and therefore it is useful to provide such analysis. Beyond this, there is an important problem of macroscopic observability of quantum phenomena in plasmas; this problem is also connected with the question of when (and which) quantum phenomena are important in quantum plasmas. The answer to this question of course depends on the models and approximations used to describe the quantum plasma. In this paper, we provide a detailed analysis of the quantum hydrodynamics model, and study the kinetic features of analytical properties of the linear dielectric response function in isotropic unmagnetized quantum plasma. In doing this, we highlight the most important, in our view, fundamental problems associated with the linear response of quantum plasma, which require further investigation.

\section{Basic methods of description of quantum plasmas \label{sec:2}}

As in the case of classical plasmas, the most complete description of quantum plasma as a system of many interacting particles is a completely hopeless task. In case of quantum plasma this task is in a sense even more hopeless than in case of classical plasma, not only because it is impossible to solve the Schr\"odinger's equation for the $N$-particle wavefunction of the system, but also because of the lack of such wavefunction for a macroscopic system that interacts, however weakly, with its environment~\cite{LL5}. Yet the problem can be significantly simplified by assuming that the plasma is nearly ideal, i.e., that the two- and higher-order correlations between its particles can be neglected. If this is the case, then the plasma can be considered as a collection of quantum particles that interact only via their collective field. As mentioned above, the most commonly used now~\cite{Manfredi_Howto} are the following models (which all in fact use the assumption of ideal plasma): (i) the quantum analog of the multistream model~\cite{Haasetal2000}, (ii) the kinetic model based on the Wigner equation for the density of quasi-probability of particle distribution in coordinates and momenta~\cite{Klim_Silin_52,Tatarskii_1983}, and, finally, (iii) the quantum hydrodynamics model. All of these models, in one way or another, are based on the Schr\"odigner's equations for the wavefunctions of plasma particles, and therefore are non-relativistic; hence they can only be used for describing non-relativistic ideal plasmas and, strictly speaking, for describing plasma oscillations with small (non-relativistic) phase velocities $\omega/k\ll c$~\cite{Melrose_Mushtaq_2009} (here $\omega$ and $k$ are frequency and wavenumber of the oscillations, respectively, and $c$ is the speed of light). We should note, however, that more general relativistic models of ``quantum plasmadynamics'' appeared recently~\cite{Melrose_Quantum_plasmadynamics}; we expect that these models will be more widely used in the future, due to their logically consistent description of both quantum particles and quantized fields. However, here we will only consider the non-relativistic models, as they are the most widely used in recent literature on quantum plasmas, probably owing to their relative simplicity.

Description of quantum plasma should be started with the models of Hartree and Hartree-Fock, in which $N$ independent Schr\"odinger equations for $N$ plasma particles are coupled via the average self-consistent field (in Hartree-Fock's model the correction due to exchange interactions is also taken into account). The Hartree and Hartree-Fock approximations for quantum plasmas are analogous to the self-consistent field approximation in description of classical plasmas, and they form a basis [lay a foundation?] for kinetic and hydrodynamic models of quantum plasmas. Therefore it is important for further considerations to write down their main assumptions.

\bigskip\noindent
\textbf{Main assumptions of the Hartree and Hartree-Fock models:}
\begin{enumerate}
\item	Plasma particles interact only through average classical (i.e., not quantized) collective fields.
\item	Plasma is ideal, $\Gamma_q=U_{\rm int}/\epsilon_F=e^2n^{1/3}/\epsilon_F\sim(\hbar\omega_p/\epsilon_F)^2\ll 1$,
    where $\epsilon_F=(\hbar^2/2m)(3\pi^2 n)^{2/3}$ is the Fermi energy of electrons, $\omega_p=(4\pi e^2 n/m)^{1/2}$ is electron plasma frequency, and $e$, $m$ and $n$ are charge, mass, and number density of electrons. We should note that for the electron gas in metals the condition $\Gamma_q\ll 1$ is in general not satisfied: in metals we have $\Gamma_q \sim 1$.
\item	Non-relativistic approximation is used; see the discussion after Eq.~(\ref{Eq.22}).
\end{enumerate}

\bigskip\noindent
\textbf{Kinetic models of Wigner-Poisson and Wigner-Maxwell}
These models are based on the Wigner equation describing time evolution of the Wigner function~\cite{Tatarskii_1983}, which is coupled with either Poisson's equation or Maxwell's equations describing the self-consistent collective electrostatic and electromagnetic field, respectively. The Wigner function describes quasi-density of quantum particle probability distribution in coordinate-momentum phase space (we call it quasi-density because the Wigner function can attain negative values, due to noncommutativity of position and momentum operators in quantum mechanics, i.e., due to uncertainty principle). The Wigner function $f(\mathbf{q},\mathbf{P},t)$ is defined from the density matrix $\rho(\mathbf{q},\mathbf{q'},t)$ of plasma in coordinate representation as follows~\cite{Klim_Silin_52}:
\begin{equation}
f(\mathbf{q},\mathbf{P}) = \frac{1}{(2\pi)^N}\int d \vec{\tau} e^{-i \vec{\tau}\cdot\mathbf{P}}
\rho\left(\mathbf{q}-\frac{1}{2}\hbar \vec{\tau},\mathbf{q}+\frac{1}{2}\hbar \vec{\tau}\right),
\label{Eq.1}
\end{equation}
where $\mathbf{q}$ and $\mathbf{P}$ are canonically conjugated generalized coordinate and momentum, $N$ is the number of components of $\mathbf{P}$ (and/or of $\mathbf{q}$) and corresponds to the number of coordinates of a particle ($N=3$ in a 3-dimensional system). The Wigner function is normalized so that
\begin{equation}
n(\mathbf{q}) = \int f(\mathbf{q},\mathbf{P})d\mathbf{P},
\label{Eq.2}
\end{equation}
where $n(\mathbf{q})$ is the number density of plasma particles. The Wigner equation, which describes evolution of the Wigner function, follows from the evolution equation for the density matrix in coordinate representation, and has the following form~\cite{Moyal_49,Klim_Silin_52}:
\begin{eqnarray}
\frac{\partial f(\mathbf{q},\mathbf{P})}{\partial t} = \frac{1}{(2\pi)^N}\frac{i}{\hbar}\int\ldots\int d \vec{\tau}
d\mathbf{k}d \vec{\eta}d\mathbf{r} e^{i\left[ \vec{\tau}\cdot( \vec{\eta}-\mathbf{P})+\mathbf{k}\cdot(\mathbf{r}-\mathbf{q})\right]}
f(\mathbf{r}, \vec{\eta})
\nonumber\\
\times\left[H\left( \vec{\eta}+\frac{1}{2}\hbar\mathbf{k},\mathbf{r}-\frac{1}{2}\hbar \vec{\tau}\right)
- H\left( \vec{\eta}-\frac{1}{2}\hbar\mathbf{k},\mathbf{r}+\frac{1}{2}\hbar \vec{\tau}\right)\right],
\label{Eq.3}
\end{eqnarray}
where $H$ is the Hamiltonian of the system.

For a system of charged particles interacting via a self-consistant electrostatic field with a potential  $\phi(\mathbf{q})$, the Hamiltonian is $H(\mathbf{q},\mathbf{P})=\mathbf{p}^2/2m+e\phi(\mathbf{q})=H(\mathbf{q},\mathbf{p})$, where $\mathbf{p}$ is the kinetic momentum of a particle (which in this case coincides with the generalized momentum $\mathbf{P}$). For such systems, the Wigner equation~(\ref{Eq.3}) becomes
\begin{equation}
\frac{\partial f(\mathbf{q},\mathbf{p})}{\partial t} + \frac{\mathbf{p}}{m}\cdot\frac{\partial f}{\partial\mathbf{q}}
= \frac{1}{(2\pi)^N}\frac{i}{\hbar}\int d \vec{\tau}d \vec{\eta} e^{i \vec{\tau}\cdot( \vec{\eta}-\mathbf{P})}
f(\mathbf{q},\mathbf{p})\left[U\left(\mathbf{q}-\frac{1}{2}\hbar\vec{\tau }\right)
-U\left(\mathbf{q}+\frac{1}{2}\hbar \vec{\tau}\right)\right].
\label{Eq.4}
\end{equation}
Equation~(\ref{Eq.4}) coupled with the Poisson's equation for the electrostatic potential $\phi(\mathbf{q})$ (in which the density of charged particles (electrons) is defined by Eq.~(\ref{Eq.2})), describes a weakly correlated system of charged particles interacting electrostatically, and is called the Wigner-Poisson model.

For a system of spinless charged particles interacting via a self-consistent electromagnetic field, the Hamiltonian is $H(\mathbf{q},\mathbf{P})=\left(\mathbf{P}-e\mathbf{A}(\mathbf{q})/c\right)^2/2m+e\phi(\mathbf{q})$,
where $\mathbf{P}$ is the canonical momentum of a particle, and $\phi(\mathbf{q})$ and $\mathbf{A}(\mathbf{q})$ are scalar and vector potentials of the electromagnetic field, respectively. By changing variables according to $\mathbf{P}=\mathbf{p}+e\mathbf{A}(\mathbf{q})/c$, where $\mathbf{p}$ is the particle kinetic momentum, the Wigner equation~(\ref{Eq.3}) is then cast in the form (we note that the corresponding equation~(30) of Ref.~\cite{Klim_Silin_52} has typos in some of the signs)
\begin{eqnarray}
\frac{\partial f(\mathbf{q},\mathbf{p})}{\partial t} + \frac{\mathbf{p}}{m}\cdot\frac{\partial f}{\partial\mathbf{q}}
& + & e\left(\mathbf{E}+\frac{\mathbf{p}\times\mathbf{H}}{mc}\right)\cdot\frac{\partial f}{\partial\mathbf{p}}
\nonumber\\
& = & \frac{1}{(2\pi)^3}\frac{1}{m}\int d \vec{\tau}d \vec{\xi} e^{i \vec{\tau}\cdot( \vec{\xi}-\mathbf{p})}
\left\{-i\frac{e}{c}f(\mathbf{q}, \vec{\xi})\left[\left( \vec{\tau}\cdot\frac{\partial}{\partial\mathbf{q}}
\right) \left(\vec{\xi}\cdot\mathbf{A}(\mathbf{q})\right)\right.\right.
\nonumber\\
-\frac{1}{\hbar} \vec{\xi}\cdot\left(\mathbf{A}\left(\mathbf{q}+\frac{\hbar \vec{\tau}}{2}\right)\right.
&-&\left.\left.\mathbf{A}\left(\mathbf{q}-\frac{\hbar \vec{\tau}}{2}\right)\right)\right]+\frac{iem}{\hbar}
\left[\phi\left(\mathbf{q}-\frac{\hbar \vec{\tau}}{2}\right)-\phi\left(\mathbf{q}+\frac{\hbar \vec{\tau}}{2}\right)\right]f(\mathbf{q}, \vec{\xi})
\nonumber\\
&+&iem\left( \vec{\tau}\cdot\frac{\partial\phi(\mathbf{q})}{\partial\mathbf{q}}\right)f(\mathbf{q}, \vec{\xi})
-\frac{e}{2c}\left(\frac{\partial f(\mathbf{q}, \vec{\xi})}{\partial\mathbf{q}}+f(\mathbf{q}, \vec{\xi})
\frac{\partial}{\partial\mathbf{q}}\right)
\nonumber\\
&&\cdot\left(2\mathbf{A}(\mathbf{q})-\mathbf{A}\left(\mathbf{q}-\frac{\hbar \vec{\tau}}{2}\right)
-\mathbf{A}\left(\mathbf{q}+\frac{\hbar \vec{\tau}}{2}\right)\right)
\nonumber\\
&-&i\frac{e^2}{2c^2}f(\mathbf{q}, \vec{\xi})
\left(2\mathbf{A}(\mathbf{q})-\mathbf{A}\left(\mathbf{q}-\frac{\hbar \vec{\tau}}{2}\right)
-\mathbf{A}\left(\mathbf{q}+\frac{\hbar \vec{\tau}}{2}\right)\right)
\nonumber\\
&&\left.\cdot\left(\frac{\partial( \vec{\tau}\cdot\mathbf{A}(\mathbf{q}))}{\partial\mathbf{q}}
+\frac{1}{\hbar}\left[\mathbf{A}\left(\mathbf{q}-\frac{\hbar \vec{\tau}}{2}\right)
-\mathbf{A}\left(\mathbf{q}+\frac{\hbar \vec{\tau}}{2}\right)\right]\right)\right\}.
\label{Eq.5}
\end{eqnarray}
Equation~(\ref{Eq.5}), coupled with the Maxwell's equations for the self-consistent electromagnetic field and a gauge condition (e.g., the Coulomb gauge $\mathbf{\nabla}\cdot\mathbf{A}=0$), describes a weakly correlated system of charged spinless particles interacting electromagnetically, and is called the Wigner-Maxwell model.

\bigskip\noindent
\textbf{Main assumptions of Wigner-Poisson and Wigner-Maxwell models:}
\begin{enumerate}
\item	Plasma is ideal, $\Gamma_q=U_{\rm int}/\epsilon_F=e^2n^{1/3}/\epsilon_F\sim(\hbar\omega_p/\epsilon_F)^2\ll 1$.
    As noted above, this condition is not satisfied for electron gas in metals, where $\Gamma_q\sim 1$.
\item	Plasma particles interact only via average collective fields that are described by Maxwell's equations (i.e., classical electrodynamics is assumed for the fields).
\item	Collisions between quantum particles are not taken into account (the models are collisionless).
\item	Non-relativistic approximation is used; see the discussion below after~(\ref{Eq.22}).
\item	Usually, spin of particles (as well as exchange interactions) are not taken into account. However, the effect of spin can still be accounted for in the workframe of non-relativistic Wigner-Maxwell model by introducing the spin distribution function and writing the corresponding kinetic equation for this function; this has been done, for example, by Silin and Rukhadze~\cite{Silin_Rukhadze_61}.

\end{enumerate}

\bigskip\noindent
\textbf{Multistream model}

This model is based on the Hartree approximation of plasma particles interacting via self-consistent collective fields only. Plasma is considered as a collection of weakly correlated ``cold beams'' formed by groups of particles with the same momenta; these beams are assumed to interact only through their collective fields. Using linearized equations of ``cold hydrodynamics'' for each of these groups of particles (beams), their current density and the corresponding dielectric permittivity tensor are calculated. Then adding up the contributions of all groups of plasma particles with the corresponding ``weight functions'', i.e., averaging these contributions over plasma equilibrium distribution function $f_0(\mathbf{p})$, one obtains the dielectric permittivity tensor of the whole plasma (see Sec.~\ref{sec:4} for details of this procedure). This procedure is equivalent to calculating the dielectric permittivity tensor directly from the Wigner equation~(\ref{Eq.5}), since the latter is based on the same assumption of weakly correlated particles interacting only via their collective fields.

\bigskip\noindent
\textbf{Main assumptions of the multistream model:} Same as for kinetic models based on the Wigner equation (see above).

\bigskip\noindent
\textbf{Model of quantum hydrodynamics}

This model, first described in Ref.~\cite{Wilhelm_70}, is constructed similar to the multistream model (note that it can also be derived from the Wigner model~\cite{Manfredi_Haas_01}): the wavefunctions of plasma particles are represented in the form~\cite{Madelung} $\psi_\alpha(\mathbf{r},t)=a_\alpha(\mathbf{r},t)\exp(iS_\alpha(\mathbf{r},t)/\hbar)$,
where $a_\alpha(\mathbf{r},t)$ and $S_\alpha(\mathbf{r},t)$ are the real functions of space and time, $\alpha$ is the particle index. The density $n_\alpha$ and velocity $\mathbf{v}_\alpha$ of an $\alpha$-th particle are defined as $n_\alpha=|\psi_\alpha(\mathbf{r},t)|^2=a_\alpha^2(\mathbf{r},t)$,
$\mathbf{v}_\alpha=\mathbf{\nabla}S_\alpha(\mathbf{r},t)/m$. The macroscopic plasma density $n(\mathbf{r},t)=\langle n_\alpha\rangle$ and velocity $\mathbf{u}(\mathbf{r},t)=\langle \mathbf{v}_\alpha\rangle$ are introduced, where $\langle\ldots\rangle$ denotes averaging over an ensemble of plasma particles, and two equations are written for $n(\mathbf{r},t)$ and $\mathbf{u}(\mathbf{r},t)$: the equation of continuity and the equation of motion, the latter of which contains two pressure-like terms~\cite{Manfredi_Haas_01} -- the classical pressure defined as $P^{cl}=mn(\langle v_\alpha^2\rangle-\langle v_\alpha\rangle^2)$, and the quantum pressure
\[
P^{q}=\frac{\hbar^2}{2m}\langle(\mathbf{\nabla}a_\alpha)^2-a_\alpha(\mathbf{\nabla}^2a_\alpha)\rangle.
\]
In order to close the set of these two equations, the following \textbf{two assumptions} are made for $P^q$ and $P^{cl}$:
\begin{enumerate}
\item	It is assumed that wavefunctions of all plasma electrons have equal amplitudes $a_\alpha(\mathbf{r},t)=a(\mathbf{r},t)$ (which nevertheless can vary in space and time), while having different phases $S_\alpha(\mathbf{r},t)$. This assumption is in agreement with the assumption of uncorrelated plasma particles: indeed, the spatial distribution of each quantum particle, defined by the amplitude $a_\alpha(\mathbf{r},t)$, is independent of the spatial distribution of other particles in the system. The assumption $a_\alpha(\mathbf{r},t)=a(\mathbf{r},t)$ implies that the spatial distribution density of each quantum particle $n_\alpha=|a_\alpha|^2$ is proportional to the density $n$ of the whole system of particles, i.e., all particles are ``smeared'' over the whole system in the same way; in other words, the size of the wavepacket representing each particle of the system is equal to the size of the whole system of particles. (We note that this assumption does not impose any restrictions on the spatial and temporal scales of the waves that can be correctly described within this model.) This assumption implies the following relation between $P^q$ and $n$~\cite{Manfredi_Haas_01}:
    \begin{equation}
    P^{q}=\frac{\hbar^2}{2m}\left[\left(\mathbf{\nabla}\sqrt{n}\right)^2-\sqrt{n}\left(\mathbf{\nabla}^2\sqrt{n}\right)\right].
    \label{Eq.6}
    \end{equation}
\item Some equation of state is assumed, which links the ``classical'' pressure of the quantum plasma
    $P^{cl}\equiv mn(\langle v_\alpha^2\rangle-\langle v_\alpha\rangle^2)$ with the macroscopic density of plasma $n(\mathbf{r},t)\equiv\langle n_\alpha\rangle$.
\end{enumerate}

Assuming a particular equation of state for the classical pressure $P^{cl}$ imposes the corresponding limitations on the applicability of thus obtained hydrodynamics model; this is discussed in more detail below in Sec.~\ref{sec:3}. Beside the assumptions 1 and 2 above, the hydrodynamics model of quantum plasma also makes all the main assumptions of the kinetic models based on the Wigner equation (see above).

\bigskip\noindent
To summarize, we list all the \textbf{main assumptions of the hydrodynamics model of quantum plasma:}
\begin{enumerate}
\item	\label{assumption:ideal} Plasma is ideal, $\Gamma_q=U_{\rm int}/\epsilon_F=e^2n^{1/3}/\epsilon_F\sim(\hbar\omega_p/\epsilon_F)^2\ll 1$. As already noted, this condition is not satisfied for the gas of conduction electrons in metals, where $\Gamma_q\sim 1$.
\item	Plasma particles interact only via average collective fields that are described by Maxwell's equations.
\item	Collisions between quantum particles are not taken into account (the models are collisionless).
\item	Exchange interactions between plasma particles are ignored.
\item	Non-relativistic approximation is used; see the discussion below after~(\ref{Eq.22}).
\item	\label{assumption:same_amplitude} The wavefunctions of all plasma electrons $\psi_\alpha(\mathbf{r},t)=a_\alpha(\mathbf{r},t)\exp(iS_\alpha(\mathbf{r},t)/\hbar)$ are assumed to have equal amplitudes $a_\alpha(\mathbf{r},t)=a(\mathbf{r},t)$ (which nevertheless can vary in space and time), while differing in their phases $S_\alpha(\mathbf{r},t)$. This imposes a relation between $P^q$ and $n$ -- the ``equation of state''~(\ref{Eq.6}) for the quantum pressure $P^q$. Note that this assumption restrains the size of plasma particle wavepackets (requiring them to be equal to the size of the whole system), but does not restrain the frequencies and wavenumbers of plasma waves that can be adequately described by this model.
\item	\label{assumption:last} Some equation of state is assumed, which relates the ``classical'' plasma pressure $P^{cl}\equiv mn(\langle v_\alpha^2\rangle-\langle v_\alpha\rangle^2)$ with the macroscopic density of plasma $n(\mathbf{r},t)\equiv\langle n_\alpha\rangle$. Usually the adiabatic equation of state is postulated, $P^{cl}=P_0^{cl}(n/n_0)^3$, with $P_0^{cl}=n_0\epsilon_F$ for degenerate electrons (i.e., for $T_e\ll\epsilon_F$, where $T_e$ is the electron temperature in energy units), or with $P_0^{cl}=n_0 T_e$ for non-degenerate electrons (i.e., for $T_e\gg\epsilon_F$). From this equation of state follows the following restriction for waves that can be correctly described within such hydrodynamics model: $k\lambda_F\ll 1$ for degenerate electrons, where $\lambda_F=v_F/\sqrt{3}\omega_p$ is the Fermi-Thomas length,, $v_F=\sqrt{2\epsilon_F/m}$ is the Fermi velocity of electrons, or $k\lambda_D\ll 1$ for non-degenerate electrons, where $\lambda_D=\sqrt{T_e/2\pi e^2 n}$ is the electron Debye length (see Sec.~\ref{sec:3} below).
\end{enumerate}

\section{On applicability range of quantum hydrodynamics equations \label{sec:3}}

In some works (see the reviews by Manfredi~et~al.~\cite{Manfredietal_arXiv}, p. 26, after Eq.~(54), and by Manfredi~\cite{Manfredi_Howto}, p. 14, the discussion after Eq.~(4.30)) the following statement is made: ``it can be shown that, for distances larger than the Thomas-Fermi screening length $L_F$ ($\lambda_F$ in our notations), one can replace $n_\alpha$ with $n$'' in the quantum pressure
\begin{equation}
P^{q}=\frac{\hbar^2}{2m}\sum_\alpha p_\alpha\left[\left(\frac{\partial\sqrt{n_\alpha}}{\partial x}\right)^2
-\sqrt{n_\alpha}\frac{\partial^2\sqrt{n_\alpha}}{\partial x^2}\right].
\label{Eq.7}
\end{equation}
As a result of this replacement, the set of hydrodynamics equations becomes (in one-dimensional case)
\begin{eqnarray}
&&\frac{\partial n}{\partial t}+\frac{\partial(nu)}{\partial x}=0,
\label{Eq.8} \\
&&\frac{\partial u}{\partial t}+u\frac{\partial u}{\partial x}=\frac{e}{m}\frac{\partial\phi}{\partial x}
-\frac{1}{mn}\frac{\partial P^{cl}}{\partial x}-\frac{1}{mn}\frac{\partial P^q}{\partial x},
\label{Eq.9}
\end{eqnarray}
where $P^q=P^q(n)$ is defined by Eq.~(\ref{Eq.6}). This statement in fact implies that replacing $n_\alpha$ with $n=\langle n_\alpha\rangle$ in~(\ref{Eq.7}), which is equivalent to postulating the ``equation of state''~(\ref{Eq.6}) for the quantum pressure, is only valid at length scale large compared to $\lambda_F$. This essentially means that postulating the ``equation of state''~(\ref{Eq.6}) for the quantum pressure $P^q$ imposes the condition that the lengths of waves correctly described by Eqs~(\ref{Eq.8})--(\ref{Eq.9}) should be large compared to the Thomas-Fermi length: $k\lambda_F\ll 1$ (for degenerate electrons, when $T_e\ll\epsilon_F$). The proof of this statement in Refs~\cite{Manfredietal_arXiv} and \cite{Manfredi_Howto}) is based merely on the fact that only for $k\lambda_F\ll 1$ the hydrodynamics equations~(\ref{Eq.8})--(\ref{Eq.9}) correctly describe the dispersion of longitudinal oscillations in degenerate electron gas (this follows from comparison of the dispersion relations derived hydrodynamically and kinetically, which is done in Refs~\cite{Manfredietal_arXiv} and \cite{Manfredi_Howto}); hence the limitation $k\lambda_F\ll 1$ must be imposed somewhere during the derivation of the closed set of hydrodynamics equations~(\ref{Eq.8})--(\ref{Eq.9}) (which is correct), namely (and this is incorrect!) -- when the ``equation of state''~(\ref{Eq.6}) for the quantum pressure $P^q$ is postulated, which is equivalent to replacing $n_\alpha$ with $n$ in (\ref{Eq.7}).

It seems appropriate to clarify this issue here. Let us show that, for plasma with degenerate electrons (i.e., for $T_e\ll\epsilon_F$), the limitation $k\lambda_F\ll 1$ of quantum hydrodynamics appears not as a consequence of postulating the ``equation of state''~(\ref{Eq.6}) for the quantum pressure $P^q$, but as a consequence of postulating a particular equation of state for the classical pressure $P^{cl}=P^{cl}(n)$, namely -- the adiabatic equation of state $P^{cl}=P_0^{cl}(n/n_0)^3$, with $P_0^{cl}=n_0\epsilon_F$ (where $n_0$ is the equilibrium electron density); it is this equation of state that leads to the following dispersion of longitudinal oscillations in degenerate plasma~\cite{Shukla_Eliasson_UFN_2010}:
\begin{equation}
\omega^2=\omega_p^2+\frac{3}{5}k^2v_F^2+(1+\aleph)\frac{\hbar^2k^4}{4m^2}  \label{eq:disp_Langmuir}
\end{equation}
where $\aleph=(48/175) m^2 v_{F}^4/\hbar^2\omega_p^2$. 

Without any additional assumptions, except the assumption of ideal nonrelativistic plasma, the equations for the plasma density $n(x,t)$ and hydrodynamic velocity $u(x,t)$ in one-dimensional case have the form (\ref{Eq.8})--(\ref{Eq.9}), with the ``quantum pressure'' $P^q=P^q (n_\alpha)$ defined by Eq.~(\ref{Eq.7}), and the ``classical pressure'' $P^{cl}$ defined as $P^{cl}(\mathrm{r},t)\equiv mn(\langle v_\alpha^2\rangle-\langle v_\alpha\rangle^2)$ (it is called ``classical'' because it corresponds to the gas pressure in the classical limit $\hbar\rightarrow 0$, while the ``quantum'' pressure $P^q$, from which the Bohm diffusion term in ~(\ref{Eq.9}) appears, does not have an analog in classical plasma). As already noted, in order to close this set of equations, one needs to introduce two simplifying assumptions: (1) postulate an equation of state for the classical pressure -- a relation between $P^{cl}$ and $n$, and (2) postulate the ``equation of state''~(\ref{Eq.6}) for the quantum pressure $P^q$, which is equivalent to replacing $n_\alpha$ with $n$ in~(\ref{Eq.7}). Let us consider these two assumptions separately and see which limitations they impose on the applicability of the resulting set of hydrodynamics equations.

First, we consider the assumptions implied by postulating the equation of state for the classical pressure; to do this, we consider the classical limit of Eqs~(\ref{Eq.8})--(\ref{Eq.9}) (to exclude, for now, the Bohm diffusion term with $P^q$, which vanishes in the classical limit). We note that in classical plasma nothing prevents us from considering the Fermi-Dirac distribution of electrons as the equilibrium distribution -- in the classical plasma we are free to construct any equilibrium distribution at our will (as long as it is stable). As indicated in the textbook by Aleksandrov~et~al.~\cite{ABR}, in collisionless plasma (for which Eqs~(\ref{Eq.8})--(\ref{Eq.9}) are written) there are two cases when the pressure can be evaluated directly, and the set of hydrodynamics equations~(\ref{Eq.8})--(\ref{Eq.9}) can be closed. The first case corresponds to processes with characteristic lengths $L$ and times $\tau$, whose characteristic velocity greatly exceeds either electron thermal velocity (for Maxwellian distribution of electrons) or electron Fermi velocity (for Fermi-Dirac distribution of electrons):
\begin{equation}
\frac{L}{\tau}\sim\frac{\omega}{k}\gg\max\{v_{T},v_{F}\}.
\label{Eq.10}
\end{equation}
In this case, following~\cite{ABR}, we can completely neglect the thermal or Fermi spread of electron velocities, which then corresponds to the case of cold plasma (ions are assumed to be cold), and from~(\ref{Eq.9}) obtain the Euler equation with zero pressure, $P^{cl}=0$. Naturally, this approximation does not yield the correction to the dispersion of plasma waves due to thermal or Fermi velocity spread of electrons; in order to obtain this correction, one needs to take into account thermal or Fermi velocity spread of electrons, which is done further below.

The second case corresponds to processes for which
\begin{equation}
v_{Ti}\ll\frac{L}{\tau}\sim\frac{\omega}{k}\ll\max\{v_{T},v_{F}\}.
\label{Eq.11}
\end{equation}
In this case the effect of electron inertia is negligibly small (so that electrons have Boltzmann distribution), and by excluding the electric field from the momentum equations for electrons and ions one obtains the set of one-fluid hydrodynamics equations. These equations are suitable for description of processes such as ion sound (with the limitation (\ref{Eq.11})), but are not suitable for description of electron oscillations, for which the electron inertia is essential; we will thus not consider this case here.

Let us return to the first of the two cases mentioned above -- the case of fast processes~(\ref{Eq.10}) -- and take into account the effect of velocity spread of plasma electrons (the ions are still regarded as cold). We consider small perturbations of equilibrium that is characterized by the Fermi-Dirac distribution of electrons (emulating degenerate electron gas), so that the pressure at equilibrium is $P_0^{cl}=n_0\epsilon_F$ (dropping the factor of order of unity). We write the equation for electron energy density (the 2-nd moment of the electron distribution function $f$) in one-dimensional case as
\begin{equation}
\frac{\partial P^{cl}}{\partial t}+u \frac{\partial P^{cl}}{\partial x}
+3P^{cl}\frac{\partial u}{\partial x}+2\frac{\partial Q}{\partial x}=0,
\label{Eq.12}
\end{equation}
where $Q$ is the energy density flux defined as $Q=(m/2)\int{(v-u)^3 f dv}$. With the assumption of fast processes~(\ref{Eq.10}), the term with the energy density flux $\partial Q/\partial x$ is small compared to the term $\partial P^{cl}/\partial t$, and can be neglected (hence the condition~(\ref{Eq.10}) implies that the corresponding process is adiabatic). As a result, Eq.~(\ref{Eq.12}) together with the continuity equation~(\ref{Eq.8}) becomes
\begin{equation}
\left(\frac{\partial}{\partial t}+u\frac{\partial}{\partial x}\right)\frac{P^{cl}}{n^3}=0,
\label{Eq.13}
\end{equation}
from which follows the equation of state of electron gas for adiabatically fast processes $P^{cl}/n^3=$constant, or $P^{cl}=P_0^{cl}(n/n_0)^3$ with $P_0^{cl}\sim n_0\epsilon_F$. Substituting $P^{cl}=P_0^{cl}(n/n_0)^3$ in~(\ref{Eq.9}) (in the classical case, i.e., without the Bohm term), we obtain the momentum equation for electrons that accounts for electron velocity spread in equilibrium (unlike the approximation of cold electrons in~\cite{ABR}). It is this equation that yields the correction $\sim k^2 v_F^2$ (or $\sim k^2 v_T^2$ for non-degenerate electrons) to the dispersion of longitudinal electron oscillations.

Therefore, the momentum equation of quantum hydrodynamics~(\ref{Eq.9}) with $P^{cl}=P_0^{cl}(n/n_0)^3$, $P_0^{cl}\sim n_0\epsilon_F$, which yields the dispersion of longitudinal electron oscillations, is obtained from the kinetic theory in the approximation of adiabatically fast processes with $\omega\gg kv_{F}$. With application to electron oscillations with $\omega\sim\omega_p$, this approximation is valid for long wavelengths: $k\lambda_F\ll 1$  for degenerate electrons, or $k\lambda_D\ll 1$ for non-degenerate electrons.

Let us go back to the quantum case. In this case: (i) degeneracy is no longer a mere result of construction of electron distribution function, but is an effect of quantum statistics (due to Pauli's exclusion principle), and (ii) the following term (the Bohm term) appears in Eq.~(\ref{Eq.9}):
\[
\frac{\hbar^2}{2m}\frac{\partial}{\partial x}\sum_\alpha p_\alpha\frac{\partial^2\sqrt{n_\alpha}/\partial x^2}{\sqrt{n_\alpha}}
\]
(the remaining terms of Eq.~(\ref{Eq.9}) are the same as in the classical case, and for them the discussion of the previous paragraphs can be repeated). For the Bohm term the following assumption is made:
\begin{equation}
\frac{\hbar^2}{2m}\frac{\partial}{\partial x}\sum_\alpha p_\alpha\frac{\partial^2\sqrt{n_\alpha}/\partial x^2}{\sqrt{n_\alpha}}
=\frac{\hbar^2}{2m}\frac{\partial}{\partial x}\left(\frac{\partial^2\sqrt{n}/\partial x^2}{\sqrt{n}}\right)
\label{Eq.14}
\end{equation}
i.e., $n_\alpha$ is simply replaced by $n$ in the Bohm term (hence, the ``equation of state''~(\ref{Eq.6}) for the quantum pressure is postulated). It can be easily seen upon substituting $n_\alpha=|\psi_\alpha|^2$ with $\psi_\alpha(x,t)=a_\alpha(x,t)\exp(iS_\alpha(x,t)/\hbar)$ in the left hand side of~(\ref{Eq.14}) that the assumption~(\ref{Eq.14}) is equivalent to the assumption~\ref{assumption:same_amplitude} of quantum hydrodynamics (see the list of assumptions at the end of Sec.~\ref{sec:2}), which by itself does not impose any constraints on the lengths of waves that can be described within the framework of quantum hydrodynamics, as already noted in Sec.~\ref{sec:2}.

Therefore, the equations of quantum hydrodynamics (\ref{Eq.8})-(\ref{Eq.9}) with $P^{cl}=P_0^{cl}(n/n_0)^3$, $P_0^{cl}\sim n_0\epsilon_F$, and $P^q(n)$ defined by Eq.~(\ref{Eq.6}), used in a number of works considering longitudinal electron oscillations in ideal quantum plasmas, can be obtained from the kinetic theory in approximation of adiabatically fast processes~(\ref{Eq.10}), which for plasma oscillations in degenerate electron plasma is equivalent to the condition $k\lambda_F\ll 1$. It is important to stress that this condition occurs not as a limitation of validity of~(\ref{Eq.14}) (i.e., of the ``equation of state'' (\ref{Eq.6}) for quantum pressure), but as a consequence of the approximation of adiabatically fast processes, for which the equation of state for the classical electron gas pressure has the form $P^{cl}=P_0^{cl}(n/n_0)^3$ with $P_0^{cl}\sim n_0\epsilon_F$ (in one-dimension case). In case of longitudinal oscillations in degenerate electron plasma, the condition $k\lambda_F\ll 1$ corresponds to the Langmuir part of the spectrum $\omega=(\omega_p^2 + 3k^2 v_F^2/5)^{1/2}$, which is correctly described by quantum hydrodynamics, unlike the essentially kinetic part of the spectrum $\omega_p/v_F\ll k\ll mv_F/\hbar$ that is associated with (kinetic) resonance $\omega\approx k v_F + \hbar k^2/2m$ and is analogous to zero sound in an almost ideal Fermi gas; see Sec.~\ref{sec:5} for more details.

\section{Dielectric permittivity tensor of quantum plasma \label{sec:4}}

Linear response of quantum plasma to electromagnetic disturbance can be described by dielectric permittivity tensor $\epsilon_{ij}(\omega,\mathbf{k})$ of the plasma. It can be calculated by solving equations governing plasma dynamics in the presence of self-consistent electromagnetic field, linearized assuming that the fields and the perturbations of the medium by these fields is small. For collisionless quantum plasma, such equation is the Wigner equation (\ref{Eq.5}). The procedure of calculation of $\epsilon_{ij}(\omega,\mathbf{k})$ from Eq.~(\ref{Eq.5}) is rather cumbersome, however, one can obtain the same result in a somewhat simpler way, based on the quantum multistream model mentioned above in Sec.~\ref{sec:2}.

It is well known (see, e.g., Ref.~\cite{Melrose_book_Instabilities:Response}) that the linear dielectric permittivity tensor of classical plasma can be calculated, rather simply, using the classical multistream model. In this model, plasma is considered as a collection of uncorrelated groups of classical particles with definite momenta (i.e., cold beams), and for each of these groups the current density and the corresponding dielectric permittivity tensor of the group is calculated. The contributions of all groups of particles are then multiplied by the momentum distribution function of plasma, $f_0(\mathbf{p})$, and summed over, thus yielding the dielectric permittivity tensor of the whole plasma. This procedure is equivalent to calculating $\epsilon_{ij}(\omega,\mathbf{k})$ directly from the linearized Vlasov equation with self-consistent electromagnetic field.

Obviously, this procedure can be generalized to quantum plasmas as well, if the latter can be adequately described as a collection of weakly correlated quantum particles, i.e., if the coupling parameter $\Gamma_q\sim(\hbar\omega_p/\epsilon_F)^2$ is small. This generalization is done, for example, by Kuzelev and Rukhadze~\cite{Kuzelev_Rukhadze_99}. Starting from the set of ``cold'' quantum hydrodynamics equations for a group $\alpha$ of quantum particles with wavefunctions $\psi_\alpha(\mathbf{r},t)=a_\alpha(\mathbf{r},t) exp\left(iS_\alpha(\mathbf{r},t)/\hbar\right)$ (without accounting for spin, i.e., for spinless particles)
\begin{eqnarray}
&&\frac{\partial n_\alpha}{\partial t} + \nabla\cdot(n_\alpha \mathbf{v}_\alpha)=0, \nonumber \\
&&\frac{\partial \mathbf{v}_\alpha}{\partial t} + \left(\mathbf{v}_\alpha\cdot\nabla\right)\mathbf{v}_\alpha = \frac{e}{m}\left(\mathbf{E}+\frac{\mathbf{v\times B}}{c}\right) + \frac{\hbar^2}{4m^2} \nabla\left(\frac{1}{n_\alpha} \left[\nabla^2 n_\alpha - \frac{1}{2n_\alpha} (\nabla n_\alpha)^2 \right] \right), \nonumber \\
&&\mathbf{j}_\alpha = e n_\alpha \mathbf{v}_\alpha,\ \ \ \ \mathbf{v}_\alpha = \nabla S_\alpha - \frac{e}{c}\mathbf{A},
\label{Eq.15}
\end{eqnarray}
and linearizing them, it is easy to calculate the conductivity tensor $\sigma_{ij}^\alpha(\omega,\mathbf{k})$ of the $\alpha$-th group of particles, by expressing their current density via the electric field as $j_{\alpha i}(\omega,\mathbf{k})=\sigma_{ij}^\alpha(\omega,\mathbf{k}) E_j(\omega,\mathbf{k})$ (for the corresponding Fourier components). Tensor of dielectric permittivity of the $\alpha$-th group of particles is then obtained as
\begin{eqnarray}
\varepsilon_{ij}^\alpha(\omega,\mathbf{k}) &=& \delta_{ij} + \frac{4\pi i}{\omega}\sigma_{ij}^\alpha(\omega,\mathbf{k})
\nonumber \\
&=& \delta_{ij} - \frac{4\pi e^2 n_{0\alpha}}{m\omega^2}
\left\{\delta_{ij} + \frac{\omega-\mathbf{k\cdot v_\alpha}}{(\omega-\mathbf{k\cdot v_\alpha})^2 - \omega_k^2}
(k_i v_{\alpha j}+k_j v_{\alpha i}) + \frac{k^2v_{\alpha i}v_{\alpha j} +
\omega_k^2 \kappa_i \kappa_j}{(\omega-\mathbf{k\cdot v_\alpha})^2 - \omega_k^2}\right\},
\nonumber
\end{eqnarray}
where $n_{0\alpha}$ is the unperturbed number density of group $\alpha$ particles, $\omega_k=\hbar k^2/2m$, and $\vec{\kappa}={\mathbf{k}}/{|\mathbf{k}|}$ is the unit vector along $\mathbf{k}$. Summing up the contributions of all groups of plasma particles with corresponding unperturbed densities $n_{0\alpha}=\int{d\mathbf{p}_\alpha f_{0\alpha}(\mathbf{p}_\alpha)}$, the linear dielectric permittivity tensor of quantum plasma is obtained as
\begin{eqnarray}
\varepsilon_{ij}(\omega,\mathbf{k})=\delta_{ij}-\frac{4\pi e^2}{m\omega^2} \int d\mathbf{p} f_0(\mathbf{p})\frac{(\omega-\mathbf{k\cdot v})^2}{(\omega-\mathbf{k\cdot v})^2-\omega_k^2} \left\{\delta_{ij}+\frac{k_i v_j+k_j v_i}{\omega-\mathbf{k\cdot v}}+\frac{k^2 v_i v_j}{(\omega-\mathbf{k\cdot v})^2} \right. \nonumber \\
\left.+ \frac{\omega_k^2(\kappa_i \kappa_j-\delta_{ij})}{(\omega-\mathbf{k\cdot v})^2} \right\}.
\label{Eq.16}
\end{eqnarray}
Writing $\varepsilon_{ij}(\omega,\mathbf{k})$ of isotropic plasma in the form~\cite{ABR}
\[
\varepsilon_{ij}(\omega,\mathbf{k})=\varepsilon^{l}(\omega,\mathbf{k}) \kappa_i \kappa_j +
\varepsilon^{tr}(\omega,\mathbf{k})(\delta_{ij}-\kappa_i \kappa_j),
\]
from~(\ref{Eq.16}) we obtain the following expressions for longitudinal $\varepsilon^l$ and transverse $\varepsilon^{tr}$ dielectric permittivities of isotropic quantum plasma~\cite{Kuzelev_Rukhadze_99}:
\begin{eqnarray}
\varepsilon^l(\omega,\mathbf{k}) &=& 1+\frac{4\pi e^2}{\hbar k^2}\int{d\mathbf{p}
\frac{\hat{D}[f_0(\mathbf{p})]}{\omega-\mathbf{k\cdot v}}},
\label{Eq.17} \\
\varepsilon^{tr}(\omega,\mathbf{k}) &=& 1-\frac{\omega_p^2}{\omega^2} +
\frac{2\pi e^2}{\hbar\omega^2} \int{d\mathbf{p} \frac{v_\perp^2}{\omega-\mathbf{k\cdot v}}\hat{D}[f_0(\mathbf{p})]},
\label{Eq.18}
\end{eqnarray}
where the difference operator $\hat{D}[f_0(\mathbf{p})]$ is defined as
$\hat{D}[f_0(\mathbf{p})]=f_0(\mathbf{p}+\hbar\mathbf{k}/2)-f_0 (p-\hbar\mathbf{k}/2)$, and
$v_\perp$ is the component of the particle velocity perpendicular to the vector $\mathbf{k}$.
Eqs~(\ref{Eq.17}) and~(\ref{Eq.18}) are equivalent to the expressions obtained directly from the linearized Wigner equation~(\ref{Eq.5}) for isotropic quantum plasma~\cite{Silin_Rukhadze_61}:
\begin{eqnarray}
\varepsilon^l(\omega,\mathbf{k}) &=& 1 + \frac{4\pi e^2}{\omega\hbar k^2} \int{d\mathbf{p}
\frac{\mathbf{k\cdot v}}{\omega-\mathbf{k\cdot v}} \hat{D}[f_0(\mathbf{p})]},
\label{Eq.19} \\
\varepsilon^{tr}(\omega,\mathbf{k}) &=& 1 + \frac{2\pi e^2}{\omega^2 k^2}
\int d\mathbf{p} [\mathbf{k\times v}]^2 \left\{f_0'(\mathbf{p}) + \frac{1}{\hbar}
\frac{\hat{D}[f_0(\mathbf{p})]}{\omega-\mathbf{k\cdot v}}\right\},
\label{Eq.20}
\end{eqnarray}
where $f_0'(\mathbf{p})$ is the derivative with respect to particle energy $\epsilon=p^2/2m$. Indeed, it is easy to show that in isotropic plasma Eqs~(\ref{Eq.19}) and~(\ref{Eq.20}) coincide with Eqs~(\ref{Eq.17}) and~(\ref{Eq.18}), respectively. Besides, same expressions~(\ref{Eq.17}) and~(\ref{Eq.18}) are obtained for an isotropic quantum plasma by Klimontovich and Silin~\cite{Klim_Silin_52}, and also by Kuz'menkov and Maksimov~\cite{Kuz'menkov_Maksimov} from their Eqs~(27) and~(28) for $\varepsilon^l(\omega,\mathbf{k})$ and $\varepsilon^{tr}(\omega,\mathbf{k})$ in the limit of no exchange interactions.

Eqs~(\ref{Eq.17})-(\ref{Eq.18}) are obtained using non-relativistic model of quantum plasma, which, strictly speaking, is not applicable for description of plasma waves with relativistic phase speeds, $\omega/k\gtrsim c$, regardless of whether the plasma particle velocities are relativistic or not~\cite{Melrose_Mushtaq_2009}. To see this, consider the processes of emission or absorption of a quantum (of longitudinal or transverse wave) with frequency $\omega$ and wave vector $\mathbf{k}$ by a plasma particle with momentum $p$ and energy $\epsilon(p)$. In relativistic treatment, i.e., for $\epsilon = \sqrt{m^2 c^4 + p^2 c^2}$ (where $m$ is the rest mass of the particle, e.g., an electron), conservation of total energy and momentum in these processes
\begin{equation}
\epsilon' = \epsilon \pm \hbar\omega,\ \ \ \mathbf{p}' = \mathbf{p} \pm \hbar\mathbf{k}
\label{Eq.21}
\end{equation}
leads to the resonance condition
\begin{equation}
\omega-\mathbf{k\cdot v} \pm \frac{\hbar}{2m\gamma}\left(k^2-\frac{\omega^2}{c^2} \right)=0,
\label{Eq.22}
\end{equation}
where $\gamma=(1-v^2/c^2)^{-1/2}$. In case of nonrelativistic particle velocities, taking
$\epsilon\approx mc^2 (1+v^2/2c^2)$ и $\mathbf{p}\approx m\mathbf{v}(1+v^2/2c^2)$ in (\ref{Eq.22}),
we obtain
\begin{equation}
\omega-\mathbf{k\cdot v} \pm \frac{\hbar}{2m} \left(1-\frac{v^2}{2c^2}\right)\left(k^2-\frac{\omega^2}{c^2}\right)=0.
\label{Eq.23}
\end{equation}
However, if in nonrelativistic approximation one formally takes $\epsilon=p^2/2m$ in~(\ref{Eq.21}), then the term $\omega^2/c^2$ does not appear at all in the resonance condition~(\ref{Eq.23}), which, strictly speaking, is incorrect in case of relativistic phase velocity of a wave in resonance with the particle, i.e., for $\omega/k\gtrsim c$. And though the effect of the $\omega^2/c^2$ term turns out to be insignificant for most processes in unmagnetized plasma, it can be of crucial importance for some processes in the presence of an external magnetic field, e.g., for the process of cyclotron maser radiation~\cite{Melrose_book_Instabilities:Maser}. One should remember that, strictly speaking, the nonrelativistic approximation is justified \textit{a priori} only when the phase speeds of waves (as well as velocities of plasma particles) are nonrelativistic~\cite{Melrose_book_Instabilities:Response}. Indeed, energy exchange in wave-particle interaction, Eq.~(\ref{Eq.21}), includes the possibility that, for example, an initially nonrelativistic particle may become relativistic after interacting with a wave quantum with sufficiently large energy/momentum.

In relativistic treatment, the dielectric permittivity tensor of an isotropic electron-positron gas (in which electrons and positrons are assumed unpolarized) has a form~\cite{Melrose_Mushtaq_2009,Melrose_Quantum_plasmadynamics}:
\begin{equation}
\varepsilon_{ij}^{rel}(\omega,\mathbf{k}) = \delta_{ij} - \frac{4\pi e^2}{m\omega^2}
\int{\frac{d\mathbf{p}}{\gamma} f_0(\mathbf{p})
\frac{(\omega-\mathbf{k\cdot v})^2}{(\omega-\mathbf{k\cdot v})^2 - \Delta^2_k}
\left\{\delta_{ij}+\frac{k_i v_j+k_j v_i}{\omega-\mathbf{k\cdot v}}+
\frac{(k^2-\omega^2/c^2) v_i v_j}{(\omega-\mathbf{k\cdot v})^2}\right\}},
\label{Eq.24}
\end{equation}
where $\mathbf{p}=m\gamma\mathbf{v}$, $f_0(\mathbf{p})=2\bar{n}(\mathbf{p})/(2\pi\hbar)^3$, $\bar{n}(\mathbf{p})$ is the sum of occupation numbers for electrons and positrons, and
\begin{equation}
\Delta_k = \frac{\hbar}{2m\gamma}\left(k^2 - \frac{\omega^2}{c^2}\right).
\label{Eq.25}
\end{equation}
We note that the denominators of the integrands in the electron and positron contributions in~(\ref{Eq.24}) correspond to products of the relativistic resonance conditions~(\ref{Eq.22}) for emission and absorption of wave quanta by relativistic particles (electrons or positrons). Indeed, symmetrizing~(\ref{Eq.24}) on $\mathbf{p}$ under the condition of isotropic plasma, $f_0(-\mathbf{p})=f_0(\mathbf{p})$, and equating thus obtained denominator to zero
\[
\left[(\omega-\mathbf{k\cdot v})^2 - \Delta^2_k\right]\left[(\omega+\mathbf{k\cdot v})^2 - \Delta^2_k\right]=0,
\]
we obtain four conditions
\begin{equation}
\omega \pm \mathbf{k\cdot v} = \pm \Delta_k,
\label{Eq.26}
\end{equation}
which correspond to resonances in the processes of emission and absorption of photons or plasmons by electrons and positrons, as well as in the processes of one-photon or one-plasmon electron-positron pair creation and annihilation~\cite{Melrose_Quantum_plasmadynamics}. The longitudinal $\varepsilon_{rel}^l(\omega,\mathbf{k})$ and the transverse $\varepsilon_{rel}^{tr}(\omega,\mathbf{k})$ dielectric permittivities of an isotropic plasma, obtained from~(\ref{Eq.24}), match with the corresponding expressions obtained by Tsytovich~\cite{Tsytovich_61} (as adjusted for the typo in the signs in Ref.~\cite{Tsytovich_61}, corrected in Eqs~(9.1.12)-(9.1.13) of Ref.~\cite{Melrose_Quantum_plasmadynamics}, and also neglecting the polarization effects).

For electron plasma (without positrons) in nonrelativistic limit $c\rightarrow\infty$ (i.e., for nonrelativistic plasma with $\gamma\rightarrow 1$, and for waves with $\omega/k\ll c$), Eq.~(\ref{Eq.24}) changes into
\begin{equation}
\varepsilon_{ij}^{NR}(\omega,\mathbf{k}) = \delta_{ij}-\frac{4\pi e^2}{m\omega^2}
\int{d\mathbf{p} f_0(\mathbf{p})\frac{(\omega-\mathbf{k\cdot v})^2}{(\omega-\mathbf{k\cdot v})^2-\omega_k^2}
\left\{\delta_{ij} + \frac{k_i v_j+k_j v_i}{\omega-\mathbf{k\cdot v}} +
\frac{k^2 v_i v_j}{(\omega-\mathbf{k\cdot v})^2} \right\}},
\label{Eq.27}
\end{equation}
differing from $\varepsilon_{ij}(\omega,\mathbf{k})$ of~(\ref{Eq.16}) in the term
\begin{equation}
\frac{4\pi e^2}{m\omega^2} \int{d\mathbf{p} f_0(\mathbf{p}) \frac{\omega_k^2}{(\omega-\mathbf{k\cdot v})^2-\omega_k^2}
\left(\kappa_i \kappa_j -\delta_{ij} \right)}.
\label{Eq.28}
\end{equation}
The disagreement of the nonrelativistic limit~(\ref{Eq.27}) of Eq.~(\ref{Eq.24}) with the expression~(\ref{Eq.16}) obtained in several different ways from nonrelativistic models of quantum plasma is due to the fact that Eq.~(\ref{Eq.24}) is obtained for a gas of unpolarized particles (electrons and/or positrons) with spins $\pm 1/2$, while Eq.~(\ref{Eq.16}) is obtained from the models that do not account for spin. We should note that the term~(\ref{Eq.28}), by which Eqs~(\ref{Eq.27}) and~(\ref{Eq.16}) differ, contributes only to the transverse dielectric permittivity of plasma $\varepsilon^{tr}$, and has no effect on the longitudinal dielectric permittivity $\varepsilon^{l}$. Therefore, nonrelativistic theories not accounting for the spin are not quite correct: while correctly describing longitudinal plasma oscillations, they incorrectly describe transverse plasma modes. Note that accounting for paramagnetic effects (associated with the spin of plasma electrons) in the nonrelativistic model~\cite{Silin_Rukhadze_61} leads to disappearing of the term (\ref{Eq.28}) in (\ref{Eq.16}); as a result, thus modified Eq.~(\ref{Eq.16}) becomes exactly equal to the nonrelativistic limit (\ref{Eq.27}) of the tensor (\ref{Eq.24}). The discussed discrepancy between the nonrelativistic responses (\ref{Eq.16}) and (\ref{Eq.27}), associated with the spin of plasma particles, is another important example showing the necessity of careful consideration of all the relevant effects, however small they may seem \textit{a priori} in the nonrelativistic approximation.

To obtain the relativistic generalization of the collective linear response~(\ref{Eq.16}) of a gas of spinless charged particles~(e.g., a gas of Cooper electron pairs with zero total spins, or a gas of bosons with spin $0$), one needs to construct the corresponding relativistic theory based on the Klein-Gordon equation~(which is a special case of the Dirac equation for spinless particles, i.e., for spinors of the 1-st rank -- scalars). However, one could instead employ a phenomenological approach to establishing the form of relativistic generalization of the dielectric tensor~(\ref{Eq.16}). Note that the only difference between the wanted relativistic response of a gas of spinless charged particles (which should turn into (\ref{Eq.16}) when $c\rightarrow \infty$) and the relativistic response of a gas of unpolarized electrons is in the term that turns into (\ref{Eq.28}) in the limit $c\rightarrow \infty$. The denominator of~(\ref{Eq.28}) in the nonrelativistic limit $c\rightarrow \infty$ corresponds to the product of two resonances~(\ref{Eq.22}) for emission and absorption of a wave quantum with energy $\hbar\omega$ and momentum $\hbar\mathbf{k}$. Therefore, in relativistic case this denominator should turn into the product of the corresponding relativistic resonances~(\ref{Eq.22}), i.e., $\omega_k^2$ should turn into $\Delta^2_k$, and $(\omega-\mathbf{k\cdot v})^2-\omega_k^2$ should turn into $(\omega-\mathbf{k\cdot v})^2 - \Delta^2_k$. Hence, in relativistic theory Eq.~(\ref{Eq.28}) should turn into
\begin{equation}
\frac{4\pi e^2}{m\omega^2} \int{d\mathbf{p} f_0(\mathbf{p})
\frac{\Delta^2_k}{(\omega-\mathbf{k\cdot v})^2-\Delta^2_k} \left(\kappa_i \kappa_j-\delta_{ij} \right)}.
\label{Eq.29}
\end{equation}
Adding the term~(\ref{Eq.29}) to~(\ref{Eq.24}), we obtain the following phenomenological expression for the tensor of dielectric permittivity of relativistic quantum plasma of spinless charged particles:
\begin{eqnarray}
\varepsilon_{ij}^{rel}(\omega,\mathbf{k}) &=& \delta_{ij} - \frac{4\pi e^2}{m\omega^2}
\int\frac{d\mathbf{p}}{\gamma} f_0(\mathbf{p})
\frac{(\omega-\mathbf{k\cdot v})^2}{(\omega-\mathbf{k\cdot v})^2 - \Delta^2_k}
\nonumber \\
&\times&\left\{\delta_{ij}+\frac{k_i v_j+k_j v_i}{\omega-\mathbf{k\cdot v}}+
\frac{(k^2-\omega^2/c^2) v_i v_j}{(\omega-\mathbf{k\cdot v})^2} +
\frac{\Delta^2_k(\kappa_i \kappa_j-\delta_{ij})}{(\omega-\mathbf{k\cdot v})^2}\right\}.
\label{Eq.30}
\end{eqnarray}
It is easy to verify that (\ref{Eq.30}) turns into (\ref{Eq.16}) in the limit $c\rightarrow \infty$, as required.

As discussed above, plasma response obtained from relativistic treatment (be it the response of unpolarized electron or electron-positron gas, or the response of a gas of spinless charged particles) differs from the corresponding plasma response obtained from nonrelativistic models, among other things, by a term proportional to $\omega^2/c^2$. The effect of this term on the dispersion of longitudinal and transverse waves in nonrelativistic quantum plasma (with $\gamma=1$) is usually small in most cases (e.g., for electron gas in metals), but becomes significant at large plasma densities, when the process of pair creation by photons and/or plasmons becomes energetically allowed in plasma. In this case an additional mechanism of wave damping, associated with electron-positron pair creation, is switched on (in addition to Landau damping). This damping occurs for waves (both longitudinal and transverse) with superluminous phase velocities, and has the energy threshold~\cite{Tsytovich_61}
\begin{equation}
(\hbar\omega)^2 > 4(mc^2)^2 + (\hbar k)^2 c^2.
\label{Eq.31}
\end{equation}
It follows that, in the limit of long wavelengths (when $\omega\sim\omega_p$), damping of longitudinal and transverse waves due to creation of real electron-positron pairs becomes significant at plasma densities $n\gtrsim 10^{32}\text{ cm}^{-3}$~\cite{Tsytovich_61}. This is, however, a rather exotic case, as such large densities can exist perhaps only in the core of dense astrophysical objects, e.g., white dwarf stars. On the other hand, the effects associated with creation of virtual electron-positron pairs do not have the strict energy threshold (\ref{Eq.31}), and thus can affect the waves in plasmas with relatively low densities. They can also affect analytical properties of plasma linear response function.

\section{Quantum kinetic effects and analytic properties of linear longitudinal plasma response \label{sec:5}}

Let us now consider the essentially quantum effects that occur in the kinetic description of collective modes in quantum plasmas. For simplicity, we only consider longitudinal oscillations resulting from an initial perturbation of quantum plasma.

In case of initial perturbation $f(\mathbf{r},\mathbf{p},0)$ (here $f$ is the Wigner function, see Sec.~\ref{sec:2}) of a uniform isotropic plasma, time evolution of the electrostatic potential in plasma is given by the following expression
\begin{equation}
\phi(t,\mathbf{r}) = \frac{1}{2\pi}
\int_{-\infty+i\sigma}^{+\infty+i\sigma}{\phi_\omega(\omega,\mathbf{k}) e^{-i\omega t} d\omega},
\label{Eq.32}
\end{equation}
where the integration is carried out in the plane of complex $\omega$ along the horizontal contour located in the upper half plane, ${\rm Im}(\omega)=\sigma>0$, and
\begin{equation}
\phi_\omega(\omega,\mathbf{k})=\frac{4\pi ie}{mk^2 \varepsilon^l(\omega,\mathbf{k})}
\int_{-\infty}^{+\infty}{\frac{g(\mathbf{k},p_x)}{\omega-kp_x/m} dp_x },
\label{Eq.33}
\end{equation}
where $p_x$ is the component of particle momentum along $\mathbf{k}$,
$g(\mathbf{k},p_x)=\int g(\mathbf{k},\mathbf{p})dp_y dp_z $, $g(\mathbf{k},\mathbf{p})$ is the Fourier transform of the initial perturbation $f(\mathbf{r,p},0)$, and $\varepsilon^l(\omega,\mathbf{k})$ is the longitudinal dielectric permittivity of plasma, defined by Eq.~(\ref{Eq.17}).

To obtain the behavior of the potential $\phi$ at large times $t$, one needs to integrate over $\omega$ in (\ref{Eq.32}) along the contour formed from the initial contour by taking the limit $\sigma={\rm Im}(\omega)\rightarrow -\infty$, while preserving the analyticity of the function $\phi_\omega(\omega,\mathbf{k})$ under the integral. (In order to do so, the function $\phi_\omega(\omega,\mathbf{k})$ in its turn should be analytically continued from the region of its definition ${\rm Im}(\omega)>0$ to the region ${\rm Im}(\omega)<0$; to do this, the contours of integration over $p_x$ in the numerator and denominator of~(\ref{Eq.33}) must be displaced from the real axis ${\rm Im}(p_x)=0$ into the lower half-plane ${\rm Im}(p_x)<0$ in such a way that the pole $p_x=m\omega/k$ is passed from below~\cite{Landau_46}.) This contour of integration over $\omega$ in (\ref{Eq.32}) must pass above all the singularities of the function $\phi_\omega(\omega,\mathbf{k})$ (analytically continued to the region ${\rm Im}(\omega)<0$) that lie on or under the real axis ${\rm Re}(\omega)$~\cite{Landau_46}. In case of classical plasma, the equilibrium distribution function $f_0$ is an entire function of $p_x$ (i.e., $f_0$ has no singularities at finite $p_x$), and the analytic continuation of $\varepsilon^l(\omega,\mathbf{k})$ to the region ${\rm Im}(\omega)<0$ is also an entire function of $\omega$. The same argument applies to the analytic continuation of the function
\begin{equation}
\int_{-\infty}^{+\infty} \frac{g(\mathbf{k},p_x)}{\omega-kp_x/m} dp_x,
\label{Eq.34}
\end{equation}
in the numerator of Eq.~(\ref{Eq.33}), if the initial perturbation $g(\mathbf{k},p_x)$ is also an entire function of $p_x$. Thus, provided $f_0(p_x)$ and $g(\mathbf{k}, p_x)$ are entire functions of $p_x$, the function $\phi_\omega(\omega,\mathbf{k})$~(\ref{Eq.33}) is the ratio of two entire functions of $\omega$. Then the only singularities of $\phi_\omega(\omega,\mathbf{k})$ are the poles defined by the roots of the equation $\varepsilon^l(\omega,\mathbf{k})=0$. The contribution of these poles into the integral~(\ref{Eq.32}) completely defines the evolution of $\phi(t,\mathbf{r})$, which in this case is a superposition of oscillations, exponentially damping (or growing, in case of non-equilibrium plasma) with time.

\begin{figure}[htb]
\includegraphics[width=8cm]{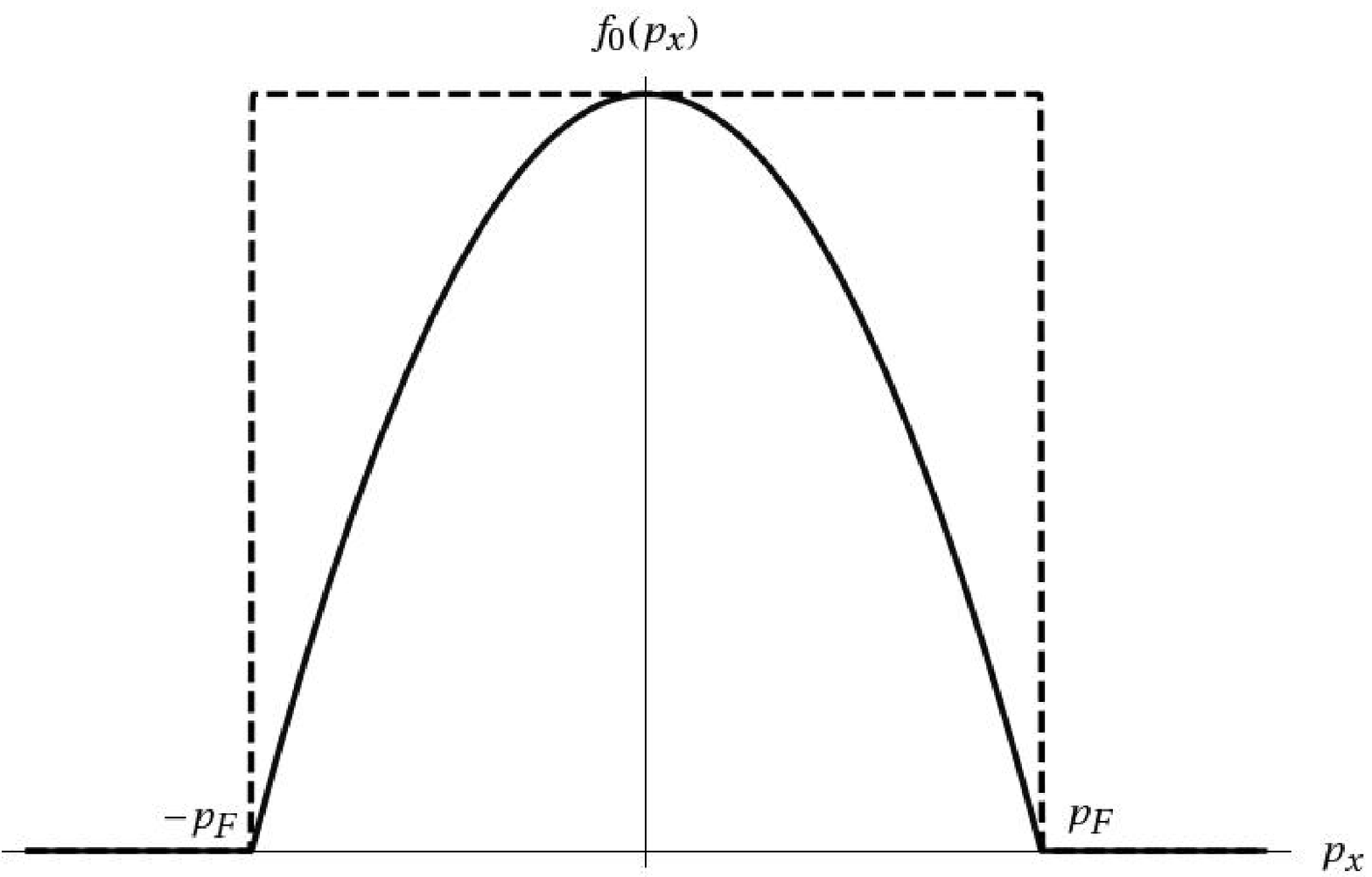}
\includegraphics[width=8cm]{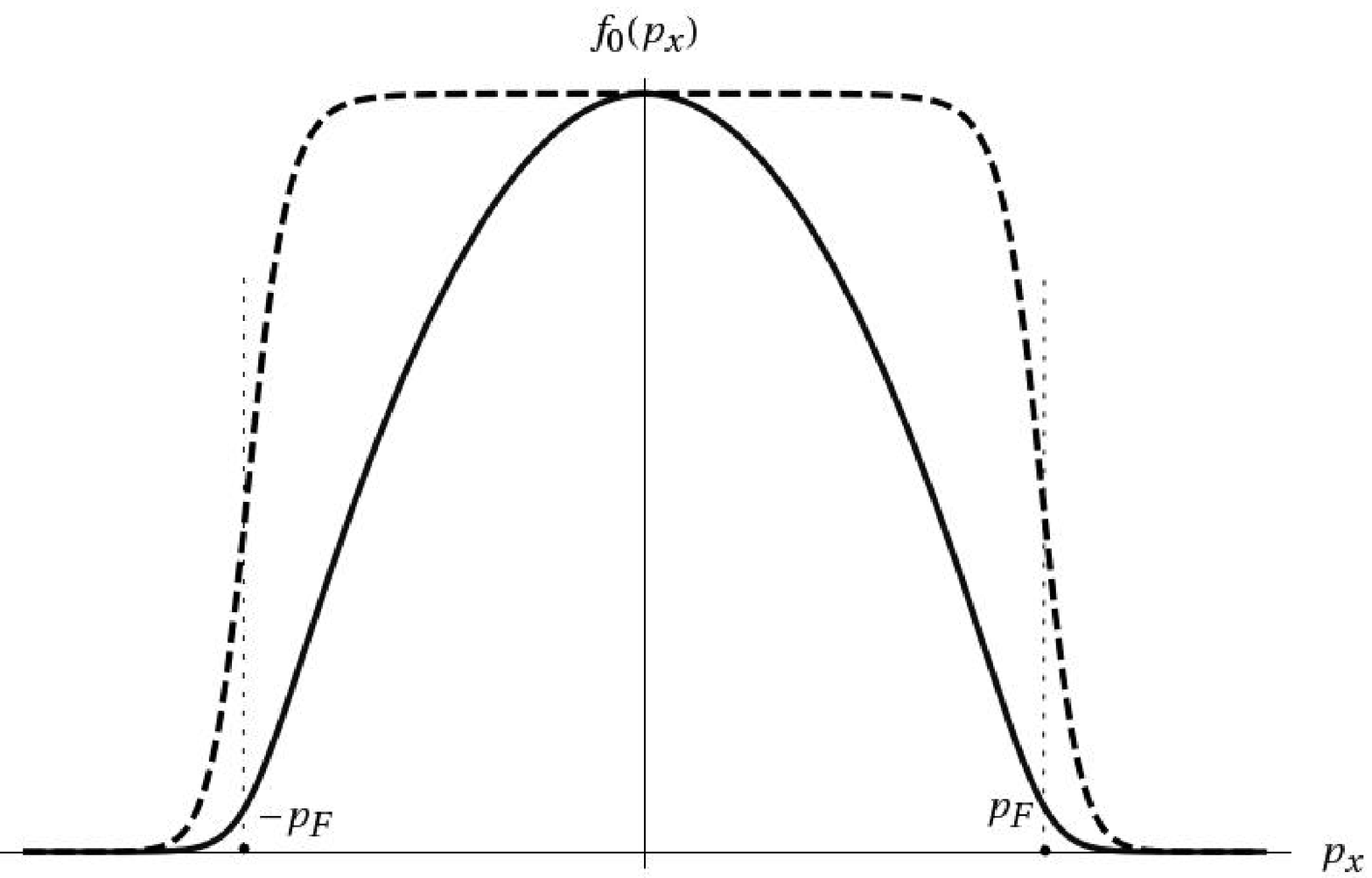}
\caption{Distribution functions $f_0^{1D}(p_x)$ for one-dimensional ($p=p_x$, shown with dashed lines) and three-dimensional ($p=\sqrt{p_x^2+p_\perp^2}$, shown with solid lines) Fermi-Dirac distribution (\ref{Eq.35}), in case of complete degeneracy ($\mu(T)/T\rightarrow\infty$, left panel) and partial degeneracy ($\mu(T)/T=10$, right panel). The functions are normalized on their corresponding values $f_0^{1D}(0)$.}
\label{Fig_f0}
\end{figure}

\begin{figure}[htb]
\includegraphics[width=9cm]{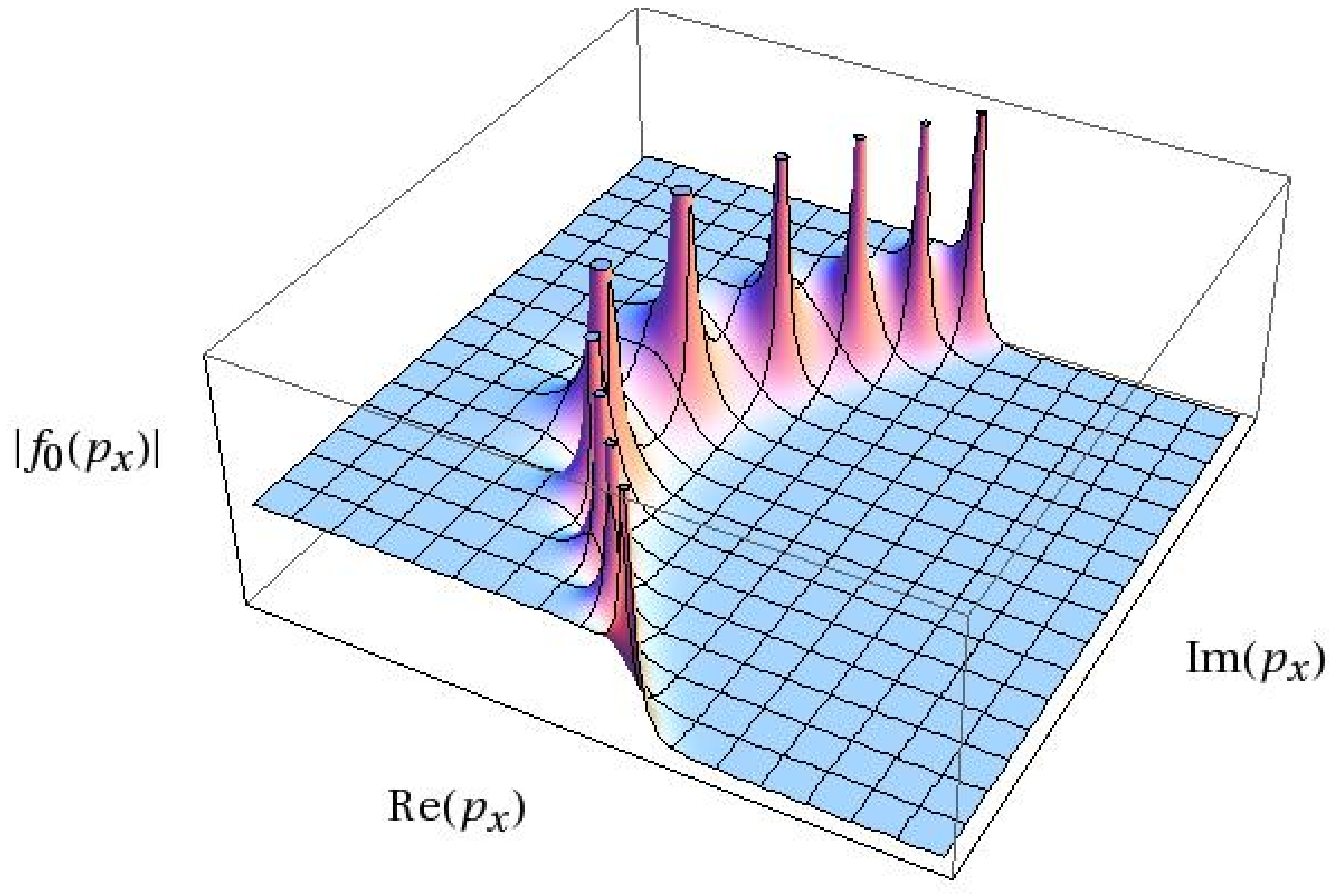}
\includegraphics[width=7cm]{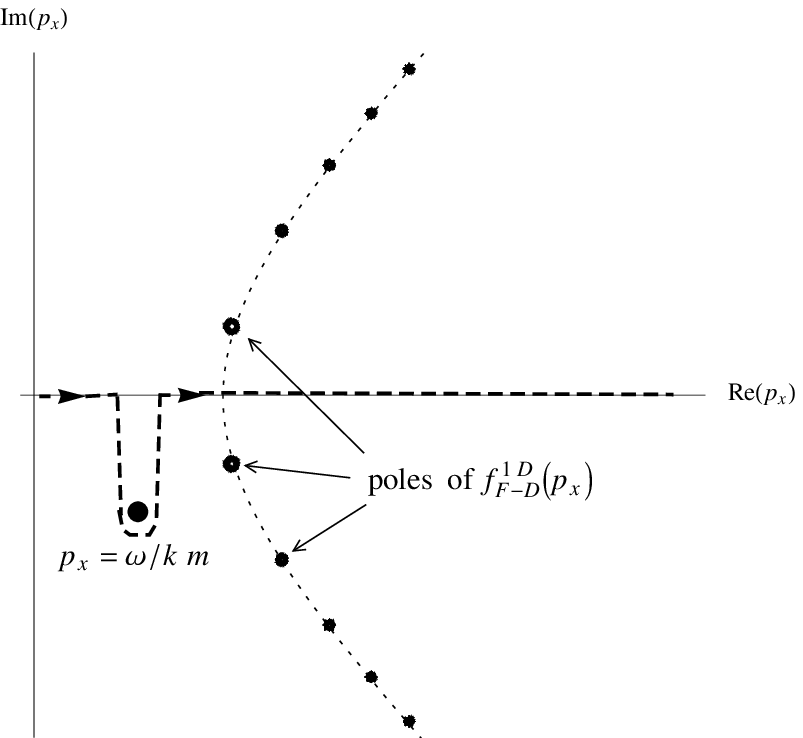}
\caption{(Color online) Left panel: absolute value of one-dimensional Fermi-Dirac distribution
$f_0(p_x)\propto\{\exp[(p_x^2/2m-\mu(T))/T]+1\}^{-1}$ as a function of complex $p_x$. Right panel: poles of one-dimensional Fermi-Dirac distribution $f_0(p_x)\propto\{\exp[(p_x^2/2m-\mu(T))/T]+1\}^{-1}$ and the contour of integration over $p_x$ (bold dashed line) for analytic continuation of $\varepsilon^l(\omega,\mathbf{k})$ into the region ${\rm Im}(\omega)<0$. The hyperbole -- the locus of the poles of $f_0(p_x)$ -- is shown by dotted line.}
\label{Fig1}
\end{figure}

However, solution of the initial value problem in quantum plasma turns out to be more complicated~\cite{Krivitskii_Vladimirov}, since the equilibrium distribution function $f_0$ is no longer an entire function of $p_x$. Indeed, an electron gas obeys Fermi statistics, and its equilibrium distribution function (Wigner function) is
\begin{equation}
f_0(p)=\frac{2}{(2\pi\hbar)^3} \left\{\exp\left[\frac{p^2/2m - \mu(T)}{T}\right]+1\right\}^{-1},
\label{Eq.35}
\end{equation}
where $T$ is the temperature and $\mu(T)$ is the chemical potential of plasma electrons. In one-dimensional case~($p=p_x$) this function (shown in Fig.~\ref{Fig_f0} by dashed lines) has singular points (poles of first order) that lie in the complex $p_x$ plane on hyperbolas intersecting the real axis ${\rm Im}(p_x)=0$ at points $\pm[2\mu(T)/m]^{1/2}$~(see Fig.~\ref{Fig1}), with the distance between adjacent singular points proportional to temperature $T$~\cite{Krivitskii_Vladimirov}, so that in the limit $T\rightarrow 0$ these points completely fill the hyperbolas. As discussed above, when performing analytic continuation of the function $\varepsilon^l(\omega,\mathbf{k})$ into the region ${\rm Im}(\omega)<0$, the contour of integration over $p_x$ in $\varepsilon^l(\omega,\mathbf{k})$ must pass the pole $p_x=m\omega/k$ from below; however, this contour should not intersect any of the poles of $f_0(p_x)$. Therefore the contour of integration over $p_x$ is ``pinched'' between the pole $p_x=m\omega/k$ (lying in the lower half-plane of complex $p_x$) and the nearest pole $p_{x0}$ of the function $f_0(p_x)$. For $m\omega/k\rightarrow p_{x0}$, i.e., when these two poles coincide, the contour unavoidably passes through (intersects) these two poles, and the function $\varepsilon^l(\omega,\mathbf{k})$ thus has singularities at $\omega_j=kp_{x0 j}/m$, where $kp_{x0 j}$ are the poles of $f_0(p_x)$. Thus in quantum plasma the function $\varepsilon^l(\omega,\mathbf{k})$ is no longer an entire function of $\omega$, and the singularities of $\varepsilon^l(\omega,\mathbf{k})$ contribute into $\phi(t,\mathbf{r})$, along with the zeros of $\varepsilon^l(\omega,\mathbf{k})$. This contribution, unlike the contribution of the zeros of $\varepsilon^l(\omega,\mathbf{k})$, might be not of the form of an exponentially damped oscillating function, but rather can be a relatively slowly (by power law) decaying function of time~\cite{Krivitskii_Vladimirov}, not necessarily oscillating.

\begin{figure}[htb]
\includegraphics[width=7cm]{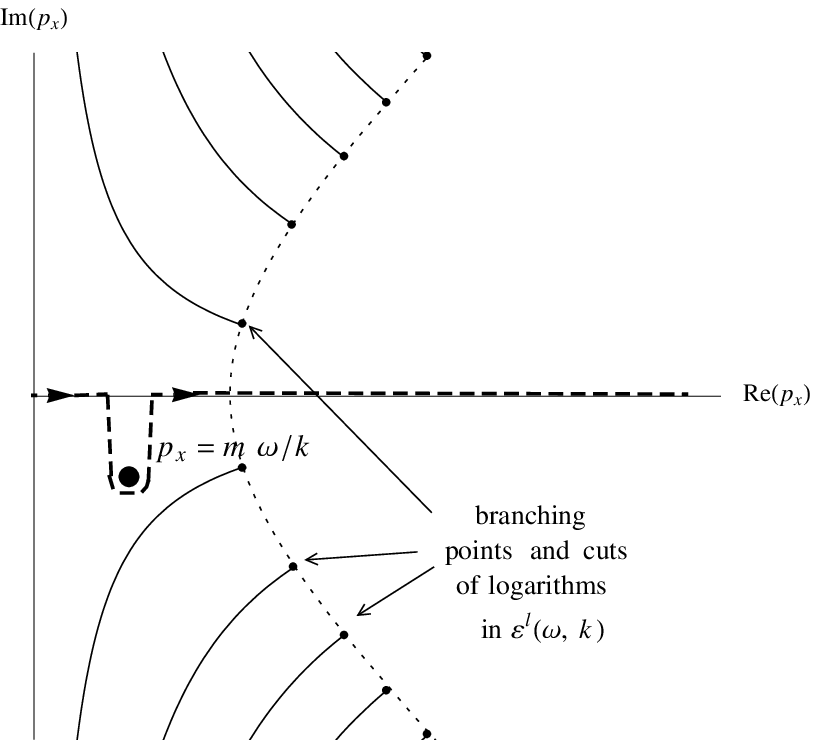}
\includegraphics[width=7cm]{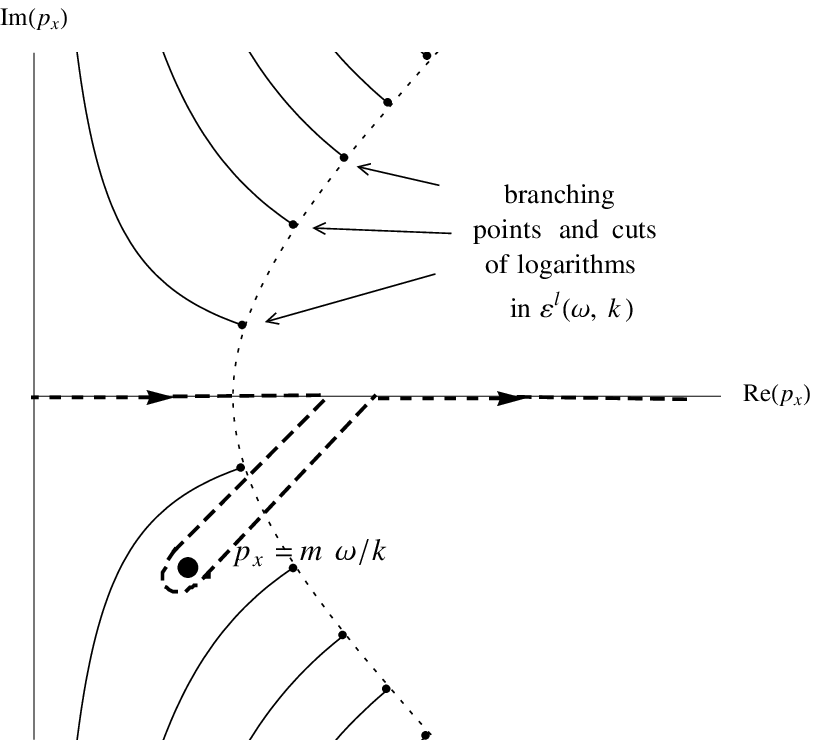}
\caption{Singularities (branching points and cuts, shown with solid lines) of the logarithmic functions under integral in~(\ref{Eq.36}), and contours of integration over $p_x$ (shown with bold dashed lines) for analytic continuation of $\varepsilon^l(\omega,\mathbf{k})$ into the region ${\rm Im}(\omega)<0$, in two cases illustrating possible location of the pole $p_x=m\omega/k$ (shown with a dot) relative to the branch cuts. The dotted line shows the locus of the branching points (the hyperbole).}
\label{Fig2}
\end{figure}

\begin{figure}[htb]
\includegraphics[width=8cm]{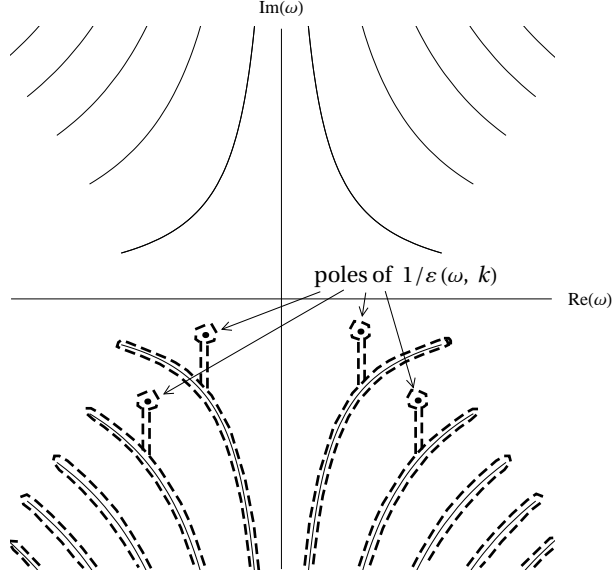}
\caption{Singularities (branching points and cuts, shown with solid lines) of
$\varepsilon^l(\omega,\mathbf{k})$ in complex $\omega$ plane, and contour of integration over
$\omega$ in~(\ref{Eq.32}), displaced into the lower half-plane to infinity and bypassing the singularities of
$\varepsilon^l(\omega,\mathbf{k})$ that lie in the lower half-plane (shown with dashed line). Both the poles of the function $1/\varepsilon^l(\omega,\mathbf{k})$ and its branch cuts in the lower half-plane of complex $\omega$ contribute to the integral in~(\ref{Eq.32}).}
\label{Fig3}
\end{figure}

However, in most cases plasma electrons are described by three-dimensional, rather than one-dimensional, distribution function, $f_0(p)\propto\{\exp[(p^2/2m-\mu(T))/T]+1\}^{-1}$, where $p$ is an absolute value of three-dimensional momentum. In this case the one-dimensional distribution function under the integral in~(\ref{Eq.17}) is defined as the three-dimensional distribution function $f_0(p)$ integrated over momenta $\mathbf{p}_\perp$ perpendicular to $\mathbf{k}$:
\[
f_0^{1D}(p_x)=\int f_0(p)d\mathbf{p}_\perp
\]
(it is shown in Fig.~\ref{Fig_f0} by solid lines), while the difference operator $\hat{D}[f_0(\mathbf{p})]=f_0(\mathbf{p}+\hbar\mathbf{k}/2)-f_0(\mathbf{p}-\hbar\mathbf{k}/2)$
acts only on $f_0^{1D}(p_x)$, $\hat{D}[f_0(\mathbf{p})]=[f_0(p_x+\hbar k/2)-f_0(p_x-\hbar k/2)] f_0(\mathbf{p}_\perp)$.
As a result, Eq.~(\ref{Eq.17}) for $\varepsilon^l(\omega,\mathbf{k})$ becomes:
\begin{eqnarray}
\varepsilon^l(\omega,\mathbf{k}) = 1 &+& \frac{4\pi e^2}{\hbar k^2} \frac{4\pi mT}{(2\pi\hbar)^3}
\int_{-\infty}^{+\infty} dp_x \frac{1}{(\omega-kp_x/m)}\left\{\ln\left[1+
\exp\left(-\frac{(p_x+\hbar k/2)^2-2m\mu(T)}{2mT}\right)\right]\right.
\nonumber \\
&-&\left.\ln\left[1+\exp\left(-\frac{(p_x-\hbar k/2)^2-2m\mu(T)}{2mT}\right) \right] \right\},
\label{Eq.36}
\end{eqnarray}
where the contour of integration over $p_x$ is again chosen to pass the pole $p_x=m\omega/k$ from below~\cite{Landau_46}.
The logarithmic functions in~(\ref{Eq.36}) are not entire functions -- each of them has singularities like branching points and branch cuts in the complex $p_x$ plane, shown in Fig~\ref{Fig2}. These singularities lead to $\varepsilon^l(\omega,\mathbf{k})$ having analogous singularities in the complex $\omega$ plane, shown in Fig.~\ref{Fig3}, which can contribute to the integral over $\omega$ in~(\ref{Eq.32}), along with the zeros of $\varepsilon^l(\omega,\mathbf{k})$. This contribution can also lead to a power-law, i.e., non-exponential character of temporal damping of an initial perturbation, similar to the case of one-dimensional Fermi distribution~\cite{Krivitskii_Vladimirov}. Note that this is a purely kinetic effect absent in quantum hydrodynamics model.

If, under some conditions, the contribution of singularities of $\varepsilon^l(\omega,\mathbf{k})$ of quantum plasma dominates over the contribution of zeros of $\varepsilon^l(\omega,\mathbf{k})$ into $\phi(t,\mathbf{r})$ at large times $t$, then the physical picture of longitudinal collective oscillation modes in quantum plasma is altered: in this case the evolution of an initial perturbation of quantum plasma at large times $t$ can be qualitatively different compared to that of classical plasma. The question of when this can happen is of fundamental interest, as it is related to macroscopic observability of quantum effects in plasma, and requires further studies. Results of this study will be published elsewhere.

\begin{figure}[htb]
\includegraphics[width=8cm]{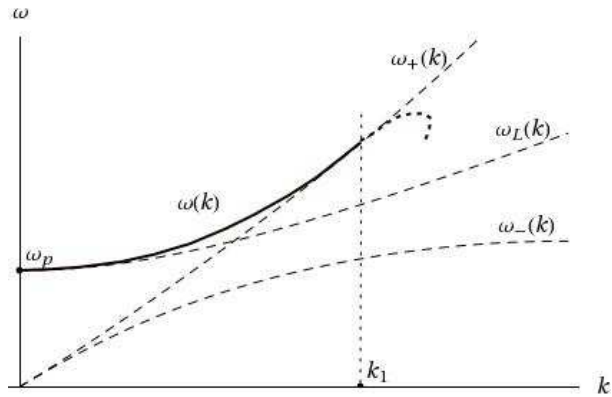}
\caption{Dispersion of longitudinal oscillations of degenerate ($T_e=0$) electron gas. The approximate solutions (\ref{Eq.37})--(\ref{Eq.38}) are shown with dashed lines. The part of the dispersion curve for which Landau damping occurs ($k>k_1$) is shown with dotted line.}
\label{Fig4}
\end{figure}

We note, however, that even in cases when the contribution from integrating along the cuts in Fig.~\ref{Fig3} is small or zero, quantum kinetic effects still significantly affect the dispersion and damping of longitudinal oscillations. In case of completely degenerate ($T_e=0$) Fermi distribution of electrons, equation $\varepsilon^l(\omega,\mathbf{k})=0$ yields the following approximate solutions at long wavelengths (note a typo in the signs in the dispersion relation (\ref{Eq.38}) of the ``kinetic mode'' in Ref.~\cite{Silin_52}):
\begin{eqnarray}
\omega_L(k) &=& \left[ \omega_p^2 + \frac{3}{5} k^2 v_F^2\right]^{1/2},\ \ \ \text{при } k\ll {\omega_p}/{v_F}, \label{Eq.37}\\
\omega_{\pm}(k) &=& (k v_F \pm \omega_k)\left\{ 1 + \exp{\left[-2-2 k^2 \lambda_F^2\right]} \right\},\ \ \ \text{при } {\omega_p}/{v_F}\ll k\ll {m v_F}/{\hbar}. \label{Eq.38}
\end{eqnarray}
These limiting solutions are shown in Fig.~\ref{Fig4}. For $k\ll {\omega_p}/{v_F}$, the dispersion of longitudinal mode (\ref{Eq.37}) corresponds to that of the usual Langmuir mode, which can also be obtained from the hydrodynamics model, see Eq.~(\ref{eq:disp_Langmuir}) of Sec.~\ref{sec:3}; thus we can call (\ref{Eq.37}) the ``hydrodynamic'' mode. Yet for ${\omega_p}/{v_F}\ll k\ll {m v_F}/{\hbar}$ the dispersion of longitudinal mode (\ref{Eq.38}) is mainly defined by the kinetic resonances $\omega-k v_F \pm \omega_k = 0$ (see Eq.~(\ref{Eq.23})) between plasmons and electrons whose velocities along $\mathbf{k}$ are equal to $v_F$. For those values of $k$ for which the ``hydrodynamic'' mode frequency~(\ref{Eq.37}) becomes close to the kinetic resonance frequency $\omega=k v_F \pm \omega_k$ (which occurs near the intersection of the dispersion curve (\ref{Eq.37}) with the resonance line $\omega=k v_F + \omega_k$), kinetic effects become dominant and ``switching'' from the hydrodynamic mode~(\ref{Eq.37}) to the kinetic mode~(\ref{Eq.38}) occurs. We also note that in degenerate plasma quantum recoil leads to an imaginary part of the longitudinal dielectric permittivity~\cite{Krivitskii_Vladimirov} for $k>k_1\approx (\omega_p/v_F)\sqrt{(3/2)(|\ln\eta|-1)}$, where $\eta=\hbar\omega_p/4\epsilon_F\sim \sqrt{\Gamma_q}$, which in turn leads to Landau damping of longitudinal oscillations with $k>k_1\approx(\omega_p/v_F)\sqrt{(3/2)(|\ln\eta|-1)}$~\cite{Krivitskii_Vladimirov} (while for $k<k_1$ Landau damping is zero). Thus quantum kinetic effects play an important (and for certain wavenumbers -- dominant) role for dispersion and damping of longitudinal collective oscillations in quantum plasmas.

\section{Summary \label{sec:6}}

In this paper we analyzed applicability limitations following from the basic assumptions of quantum plasma models most widely used in recent literature. Lack of understanding of these limitations can lead to incorrect interpretation of results obtained with a particular model. For example, description of longitudinal oscillations of degenerate electron gas with the model of quantum hydrodynamics yields a dispersion relation that is only valid for $k\lambda_F\ll 1$, which is not always stated explicitly in the literature. Moreover, the very reason for this limitation on lengths of waves described by quantum hydrodynamics is not clearly understood by all. In Sec.~\ref{sec:3} we showed that the limitation $k\lambda_F\ll 1$ appears in quantum hydrodynamics as a result of postulating a particular (adiabatic) equation of state for the ``classical'' pressure, and not as a result of postulating (\ref{Eq.6}) for the ``quantum'' pressure.

We also discussed the linear response of quantum plasma, and pointed out that it is conceptually incorrect to use the response obtained from nonrelativistic models of plasma when describing waves with relativistic phase speeds $\omega/k\gtrsim c$, regardless of whether the plasma itself is relativistic or not. Besides, we pointed out that, even for unmagnetized plasmas, ignoring the effect of spin leads to incorrect dielectric permittivity tensor, which in particular leads to incorrect dispersion relation for transverse waves.

Finally, we discussed quantum kinetic effects associated with nontrivial analytic properties of the complex linear plasma response function, occurring both due to quantum degeneracy of electron distribution and due to quantum recoil. These effects are hardly discussed in the literature, with rare exception, while being of fundamental interest. In particular, correct account of the analytic properties of quantum plasma linear response function can significantly change the physical picture of evolution of collective oscillations in both unbounded and bounded quantum plasma (Landau's initial value and boundary value problems~\cite{Landau_46}).

{\bf Acknowledgments}
The authors thank D.~Melrose and R.~Kompaneets for useful discussions in the course of writing this note.
This work was supported by Australian Research Council~(ARC).


\end{document}